\documentclass[a4,epsf,8pt]{revtex4}

\usepackage{amsmath}
\usepackage{amssymb}
\usepackage{graphicx}
\usepackage{color}
\begin{document}

\newcommand{\beq}{\begin{equation}}
\newcommand{\eeq}{\end{equation}}
\newcommand{\beqa}{\begin{eqnarray}}
\newcommand{\eeqa}{\end{eqnarray}}
\newcommand{\bmat}{\begin{displaymath}}
\newcommand{\emat}{\end{displaymath}}

\newcommand{\eq}[1]{Eq.~(\ref{#1})}

\newcommand{\lan}{\langle}
\newcommand{\ran}{\rangle}

\newcommand{\tav}[1]{\left\lan #1 \right\ran}
\newcommand{\sav}[1]{\left [\hspace*{-.13cm}\left| #1 \right|\hspace*{-.13cm}\right ]}
\newcommand{\lsav}{\Biggl [\hspace*{-.13cm}\Biggl | }
\newcommand{\rsav}{\Biggr |\hspace*{-.13cm}\Biggr ]_{\rm av} }

\newcommand{\blue}[1]{\textcolor{blue}{#1}}

\title{Vortex jamming in superconductors and granular rheology}


\author{Hajime Yoshino$^{1}$, Tomoaki Nogawa$^{2}$ and Bongsoo Kim$^{3}$}
\address{$^1$Department of Earth and Space Science, Faculty of Science,
 Osaka University, Toyonaka 560-0043, Japan\\
$^2$Division of Physics, Hokkaido University,
Sapporo, Hokkaido 060-0810 Japan\\
$^3$Department of Physics, Changwon National University, 
Changwon 641-773, Korea 
}

\begin{abstract}
We demonstrate that a highly frustrated anisotropic Josephson junction array(JJA) on a square lattice exhibits a zero-temperature jamming transition, which shares much in common with those in granular systems.
Anisotropy of the Josephson couplings along the horizontal and vertical directions plays roles similar to normal load or density in granular systems.
We studied numerically static and dynamic response of the system 
against shear, i.~e. injection of external electric current  at zero
 temperature. 
Current-voltage curves at various strength of the anisotropy 
exhibit universal scaling features around the
jamming point much as do the flow curves in granular rheology, 
shear-stress vs shear-rate. It turns out that at zero temperature the jamming transition occurs
 right at the isotropic coupling and
anisotropic JJA behaves as an exotic {\it fragile vortex matter}: 
it behaves as superconductor (vortex glass) into one direction 
while normal conductor (vortex liquid) 
into the other direction even at zero temperature.
Furthermore we find a variant of the theoretical model for
the anisotropic JJA quantitatively 
reproduces universal master flow-curves of the granular systems.
Our results suggest an unexpected common paradigm stretching over seemingly
unrelated fields - the rheology of soft materials and superconductivity.
\end{abstract}

\maketitle

Physics continues to thrive on analogy \cite{foster}.
Rheological properties of matters \cite{rheologybook} and
electric transport properties of superconductors \cite{Tinkam}
exhibit intriguing analogies.
The flow curves in rheology, the shear-stress vs the shear-rate,
correspond to the current-voltage curves in superconductors \cite{Yoshino}. 
Exploring this analogy further we here demonstrate by computer simulations 
that there exist the zero-temperature jamming transitions and glassy non-linear rheology,
originally found in granular and other materials
\cite{Cate-Wittmer-Bouchaud-Claudin,Liu-Nagel,Jamming-Book,Nagel-group,saclay,durian-group,Weeks,Otsuki-Sasa,Hatano-Otsuki-Sasa,Olsson-Teitel,Xu-Hern,Hatano},
in a class of highly frustrated anisotropic Josephson junctions arrays
(JJA) on a square lattice. Our key observation is that anisotropy 
of Josephson coupling plays the role of normal load or density in granular system such that 
a jamming transition takes place in the limit of isotropic Josephson coupling at zero temperature. 
Combined with accumulating evidences that the  (vortex) liquid-glass 
transition occurs at zero temperature in {\it isotropic} JJA, 
our result provides a strong evidence that the isotropic coupling point at zero
temperature is an ideal example of the so called J-point (Jamming point) \cite{Liu-Nagel} 
and that the anisotropic
JJA is a promising system which allows explorations of both (athermal)
unjamming-jamming transition and (thermal)
liquid-glass transitions in a unified manner 
in a {\it single system} as originally proposed
in the context of granular and glassy materials \cite{Liu-Nagel}.
Furthermore, we show that a variant of the original JJA model 
emphasizing elastic nature in a particular direction 
can quantitatively reproduce scaling features 
of granular jamming transitions observed near the J-point.

\begin{figure}[h]
\includegraphics[width=0.6\textwidth]{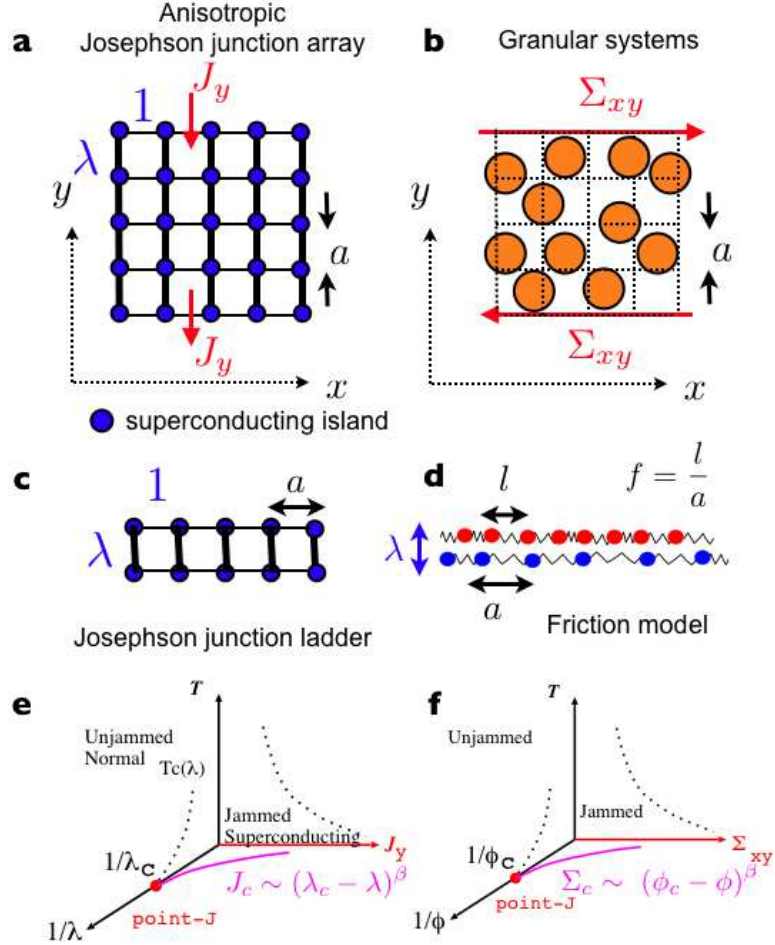}
\caption{ {\bf Highly frustrated anisotropic JJA and granular materials.}
We consider an array of Josephson junctions on a square lattice 
with linear size $L$ (in the unit of lattice spacing $a$) (a).
The ladder version of the JJA model (c) can be 
mapped onto a friction model (d).
The control parameter corresponding to the number density of granular 
particles $\phi$ or normal load (f) 
is the strength of the anisotropy of the Josephson 
coupling $\lambda$ (e). 
}
\end{figure}

The JJA \cite{Tinkam,Teitel-Jayaprakash,Mooji-group} 
is a network of superconducting islands as depicted in Fig.~1 (a). 
The phase of superconducting order parameter of the islands at site $i$,
$\theta_{i}$, 
is coupled to its nearest neighbors by the Josephson junctions.
External transverse magnetic field $B$ thread the
cells in the forms of flux lines each of which carrying
a flux quanta $\phi_{0}$. On average each unit cell of the square lattice
carries $f= B a^{2}/\phi_{0}$ flux lines.
A flux line threading a unit cell induces a vortex of the phases 
around the cell much as a dislocation in a crystal.
Here $A_{ij}$ is the vector potential. 
The static and dynamic properties of the JJA under transverse magnetic field
are known to be described to a good accuracy by the energy term
associated with the Josephson couplings \cite{Tinkam,Teitel-Jayaprakash},
\beq
H= \sum_{<i,j>} K_{ij} E_{ij}(\theta_{i}-\theta_{j}-A_{ij})
\qquad E_{ij}(\psi)=-\cos(\psi)
\label{eq-jja-hamiltonian}
\eeq
where $K_{ij}$ is the strength of the Josephson coupling. 

In mid 80's a tantalizing possibility of a {\it superconducting glass} 
in the JJA has been
raised by Halsey \cite{Halsey}: if the transverse magnetic field is tuned in such a way that
the number density $f$ of the vorticies takes {\it irrational} values, a glassy state may be
realized at low temperatures because the vorticies may not be able to form 
periodic structures called vortex lattices which are analogous to
ordered structures of dislocations in Frank-Kasper phases.
It has been argued that the frustration due to the gauge field like $A_{ij}$ 
in \eq{eq-jja-hamiltonian}
mimics geometric frustration in structural glasses (see \cite{Tarjus-review} for a review on the perspective on the
frustration-based point of view on the glass transition).
Indeed equilibrium relaxations were similar to the primary relaxation
observed in typical fragile supercooled liquids \cite{Lee-Kim,Lee-Kim-Lee}.
Now recent studies 
appear to convincingly suggest that the putative (thermal) liquid-glass
transition is actually taking place only at zero temperature exhibiting 
diverging length scale(s) \cite{Park-Choi-Kim-Jeon-Chung,Granato,Granato2}. 
We note that all these observations are made on systems
in which the Josephson couplings $K_{ij}$ are {\it isotropic}.

As stated at the beginning our key observation is that 
the anisotropy of the Josephson coupling 
$K_{ij}$ is a relevant variable and plays the role similar to the particle 
normal load or density 
in the jamming of granular materials \cite{Jamming-Book} 
(Fig.~1 (a)(b)). We parametrize the anisotropic coupling as,
\beq
K_{ij} = \left \{ \begin{array}{cc}
1 & \mbox{for $ \langle i,j \rangle$ directed along the $x$ axis}\\ 
\lambda & \mbox{for $\langle i,j \rangle$ directed along the $y$ axis}\\ 
\end{array} \right.
\label{eq-kij}
\eeq
Such an anisotropic JJA can be realized experimentally
by controlling the width of the junctions \cite{anisotropic-JJA}.
Let us also note that similar anisotropic couplings arise naturally
also in cuprate high-Tc superconductors \cite{Hu-Tachiki}
and charge-density-wave systems \cite{Nogawa-Nemoto}.

Another key observation is that the model \eq{eq-jja-hamiltonian}
can be slightly modified to build an effective Eulerian model 
for rheology of granular systems under horizontal shear as shown 
in Fig.~1 (b). We assume that particles move predominantly parallel to 
the direction of driving ($x$ axis) confined within the horizontal
layers without passing over others in the same layer.
Then the variable $\theta_{i}$ can be interpreted as
a {\it phase variable} \cite{Yoshino} by which the number 
density of particles within the $i$-th cell is described, for instance,
as $\rho_i=(1/2)(1+\cos(\theta_{i}))$. Here the size $a$ of the cell 
corresponds to the typical scale of particles and their mutual distances
\cite{continuous-limt}. The sinusoidal intra-layer couplings 
$E_{ij}(\psi)=-\cos(\psi)$ in \eq{eq-jja-hamiltonian}
are replaced by elastic couplings $E_{ij}(\psi)=\psi^{2}/2$
while the sinusoidal form is kept for the inter-layer couplings to
allow phase slips between different layers.
Let us call such a model as {\it semi-elastic model}.
Here the vorticies represent dislocations.
We assume that the geometrical frustration induced
by the gauge field $A_{ij}$
mimics real frustrations in granular 
and other glassy systems  \cite{Tarjus-review}.
In the absence of the frustration ($A_{ij}=0$) the semi-elastic model 
exhibits a non-linear rheology associated with 
a Kosterlitz-Thouless transition at finite temperatures \cite{Yoshino}.

The dynamics of the models can be described by
the equation of motion,
\beq
\frac{d\theta_{i}}{dt}=v_{i}, \qquad \qquad
m \frac{d v_{i}}{dt}=-\frac{\partial H}{\partial \theta_{i}}+F_{i}
\label{eq-motion}
\eeq
Here the frictional force $F_{i}$ is given by
$F_{i}=-\gamma \sum_{j}(v_{i}-v_{j})$ with the summation taken over
the $4$ nearest neighbours of $i$.  
This equation of motion is nothing but the standard resistively 
and capacitively shunted junction  (RCSJ) dynamics \cite{Tinkam,Granato},
which can also be viewed as a model for rheology 
of the layered systems \cite{Yoshino}.
For simplicity we choose mass (capacitance) $m=1$,
damping constant (resistance) $\gamma=1$. 
Here we focus only on the zero temperature dynamics. 
Appropriate thermal noise can be added to \eq{eq-motion} 
for finite temperature dynamics. 
For the semi-elastic model  we assumed
two different types of constitutive
relations for the frictional force $F_{i}$ in \eq{eq-motion}: 
(1)  Newtonian viscous friction $F_{i}=-\sum_{j}(v_{i}-v_{j})$ 
and (2) Bagnold's friction \cite{Bagnold,Mitarai-Nakanishi}  $F_{i}=-\sum_{j}|v_{i}-v_{j}|(v_{i}-v_{j})$ 
where the sum is taken over the nearest neighbours on the two adjacent layers.

The anisotropic coupling \eq{eq-kij} can be motivated by recalling
a well known problem in the science of friction.
A class of friction models related to the Frenkel-Kontorova model \cite{FK-review} 
is known to exhibit the so called Aubry's transition \cite{aubry} 
which is a kind of jamming transition at zero temperature.
In Fig.~1 (d) we display a friction model proposed by Matsukawa and Fukuyama
 \cite{matsukawa-fukuyama} which consists of two layers of atoms 
representing surfaces of two different solids. 
In general the ratio (winding number) $f=l/a$ of the mean atomic spacings 
on the two different materials takes irrational values.
The atoms in the same layers are connected to each other by springs while
those on different layers interact with each other via short-ranged 
interactions of strength $\lambda$, which mimic the normal load.
In the weak coupling regime $\lambda < \lambda_{c}$, the two chains of atoms
slide smoothly with respect to each other thanks to the incommensurability.
On the other hand in the strong coupling regime $\lambda > \lambda_{c}$ 
the system is pinned into amorphous metastable states
and a finite static frictional force or yield stress emerges. Note that
an anisotropic JJ ladder \cite{Denniston-Tang-ladder} 
shown in Fig.~1 (c) which consist of two horizontal layers 
can be viewed as an Eulerian formulation of the friction model. 
The irrational winding number $f$ can be identified with
the irrational number density $f$ of vorticies in a unit cell 
in the JJ ladder.

An important consequence of the anisotropic coupling \eq{eq-kij}
in 2 (and higher) dimensions
is that the effective repulsive long-ranged interactions between 
the vorticies become anisotropic. For $\lambda <1$ the vorticies
will tend to align vertically since the repulsive force is stronger along
the $x$ axis, which make it  much harder 
for the vorticies to move along the $y$ axis, i.e. the
direction with weaker coupling.
Of course the situation becomes reversed for $\lambda >1$.

The rigidity of the system can be probed 
by applying an external current just as external shear stress is applied
on a solid. What corresponds to the shear stress $\Sigma_{xy}$ along 
the $x$ axis (Fig.~1 (b)) is the vertical external 
electric current $J_{y}$ (Fig.~1 (a)) \cite{Yoshino}. 
Then vorticies (dislocations) 
are driven along $x$ axis by the Lorentz force.
The resultant electric field $E_{y}$ which is proportional to
the average velocity of the vorticies corresponds to 
the shear-rate $\dot{\Gamma}_{xy}$ which measures the rate of plastic
deformations in rheology. If the vorticies don't move significantly 
resisting against the Lorentz force, the energy dissipation is 
negligible and the system remains macroscopically superconducting.
In practice, we apply shear to the system by forcing the top 
and bottom layers (walls) to 
move along the opposite directions at constant velocities. 
We measure the resultant electric field $E_{y}$
(shear-rate $\dot{\Gamma}_{xy}$) defined as the slope of phase 
velocity $v_{i}$ developed in the system along the $y$ axis. 
Electric current flowing through a junction from site $i$ to $j$ 
is defined as $\sin(\theta_{i}-\theta_{j}-A_{ij})$.
The currents running through the junctions
parallel to $x$ and $y$ axes correspond
to the shear stresses $\Sigma_{xx}$ and $\Sigma_{xy}$ respectively in rheology.

By construction of the system, static and dynamic resposes to $J_{x}$
at anisotropy $\lambda$ at $T=0$ is just the same as those to
$J_{y}$ with $\lambda'=1/\lambda$. Thus in the following we only display results of 
resposes  to $J_{y}$.

We numerically solved the equation of motion  \eq{eq-motion} by the 4th order Runge-Kutta method
\cite{runge-kutta}. Periodic boundary condition is imposed along the $x$ axis only. 
For a given irrational  vortex density $f$ we used its rational 
approximations $p/q$ with integer $p$ and $q$
in systems of sizes $L=n q$  (with $n=1$ or $2$).
To explore larger length/time scales we use systematically better 
approximants to prevent commensurability (or matching) effects. (See
APPENDIX \ref{appendix-rational-approximants}) 
Before starting measurements, we checked that the velocity profile 
becomes linear into the $y$ axis without shear-bands and that observed quantities 
do not depend on the prior shear histories.

\begin{figure}[h]
\includegraphics[width=0.7\textwidth]{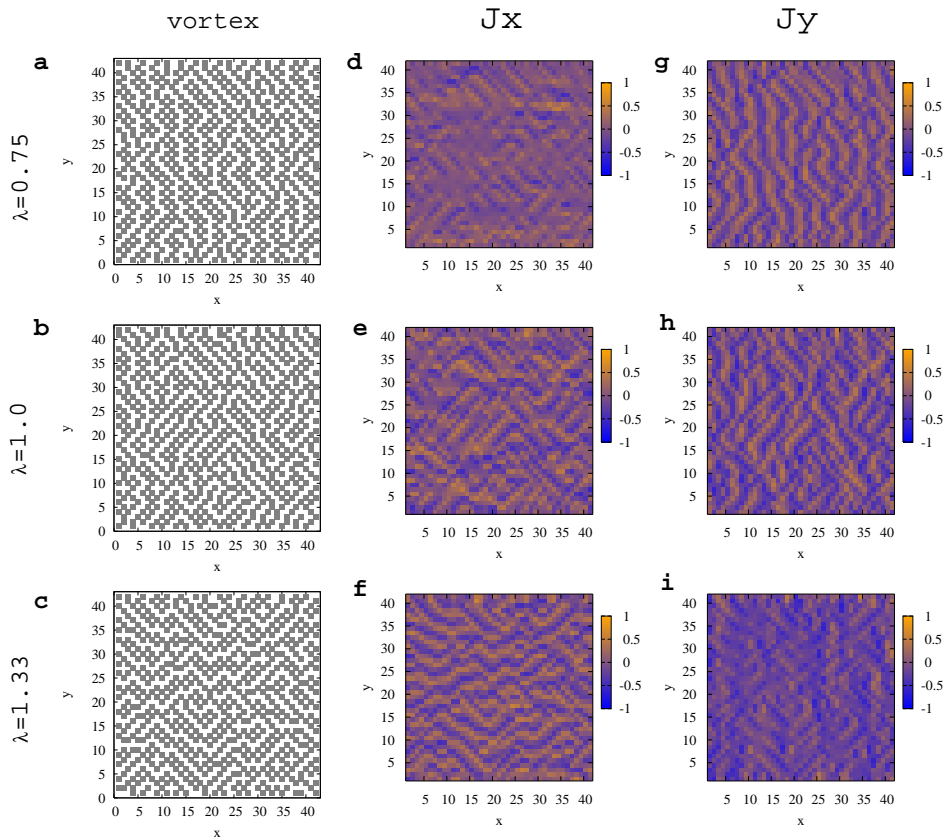}
\caption{ {\bf Snapshots of vorticies and local electric-currents 
in the steady states under shear at zero temperature.} 
Vorticies (a-c) tend to align into the direction with weaker coupling.
Local electric currents (d-i) exhibit 
chain-like configurations which are reminiscent of the ``force chains''
observed in granular materials. 
The ``current-chains'' tend to percolate into the 
direction with weaker coupling.
The system size is $L=42$ with $f=8/21$ which approximates $(3-\sqrt{5})/2=0.3819..$.
The system is sheared
such that  the electric field $E_{y}$ (shear rate) is $0.006$. 
}
\end{figure}

Shown in Fig.~2 are snapshots of the vorticies
and local currents under shear of the anisotropic JJA. Chains of electric 
currents reminiscent of ``force chains'' \cite{Cate-Wittmer-Bouchaud-Claudin,Jamming-Book} 
in granular materials can be noticed. The configurations at the isotropic point
$\lambda=1$ appear to manifest
the diagonal stripe structures found in the ground states \cite{halsey-staircase,gupta,Denniston-Tang}.
As expected the vorticies and the electric currents
flowing along the trains of vorticies
tend to align into the direction with weaker coupling.
This observation strongly suggests that a jamming transition takes place
at the isotropic point $\lambda_{c}=1$. 
The system behaves as a fluid (unjammed phase) for
$\lambda < \lambda_{c}$ and amorphous solid (jammed phase) 
for  $\lambda > \lambda_{c}$ with respect to $J_{y}$
as depicted in Fig.~1 (e). 
Furthermore it is interesting to note that the jammed 
state is inevitably fragile 
in somewhat similar sense as proposed 
in the context of granular matters \cite{Cate-Wittmer-Bouchaud-Claudin}:
the system with a given $\lambda$ can resist against shear only 
into one direction (i.~e. superconducting). 
Thus we may call such a state of matter as {\it fragile vortex matter}
in the same spirit of \cite{Cate-Wittmer-Bouchaud-Claudin}.
The qualitative features are essentially the same in the semi-elastic model
except that vorticies move only into $x$ axis in the latter model. 

\begin{figure*}[h]
\includegraphics[width=0.45\textwidth]{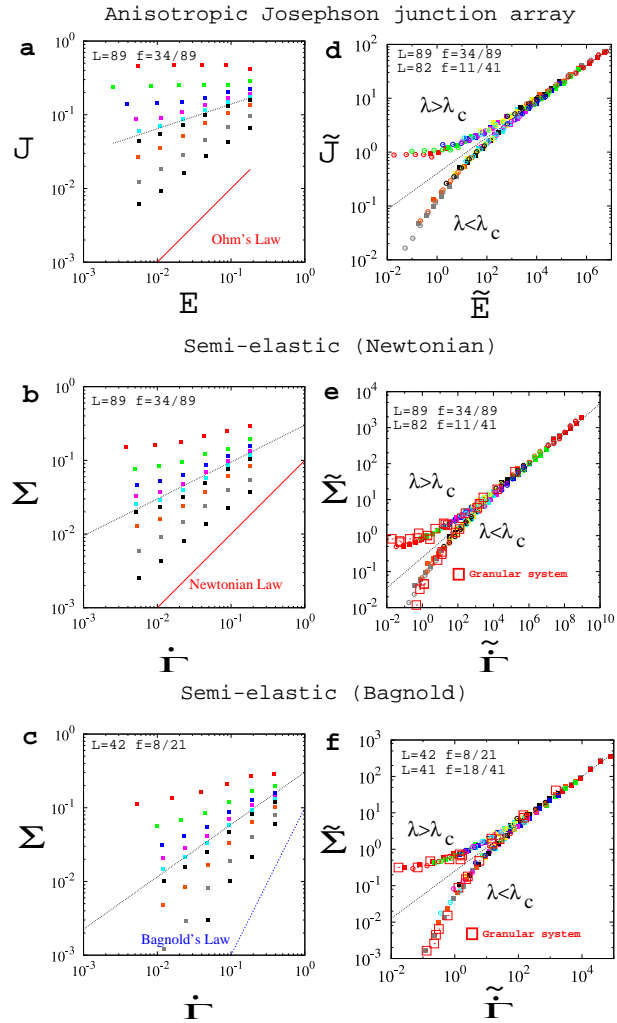}
\caption{ {\bf Current-voltage curves 
and flow curves.} 
The left panels (a,b,c) show the current-voltage
of the JJA (a), flow curves
of the semi-elastic model with Newtonian friction (b)
and that with Bagnold friction (c).
The right panels (d,e,f) are corresponding scaling plots.
The master-curves obtained in numerical simulations of 
2 dimensional granular systems \cite{Hatano-Yoshino}
with coefficient of restitution equal to/smaller than $1$
are included in (e) and (f)  respectively.
In the right panels (a,b,c) the strength of the inter-layer coupling $\lambda$ is varied
as $\lambda=0.625,0.75,0.875,0.95,1.0,1.050,1.125$ and $1.5$
from the bottom to the top curves. The dotted lines are
power law fits to the curve at $\lambda_{c}=1$ which yields
$1-\alpha=\beta/\Delta=0.34(3)$ (JJA),$0.43(6)$ (semi-elastic-Newtonian)
and $0.65(4)$ (semi-elastic-Bagnold). 
The scaling plots in the right panels (d,e,f) show $\tilde{J} \propto  J/(\lambda-\lambda_{c})^{\beta}$ vs
$\tilde{E} \propto E/(\lambda-\lambda_{c})^{\Delta}$ and
$\tilde{\Sigma} \propto \Sigma/(\lambda-\lambda_{c})^{\beta}$
vs $\tilde{\dot{\Gamma}} \propto
 \dot{\Gamma}/(\lambda-\lambda_{c})^{\Delta}$ with $\lambda_{c}=1$. 
The resultant values of the exponent $\Delta$ which give best scaling 
collapse are $3.5$ (JJA),$4.2$ (semi-elastic-Newtonian)and $2.4$
(semi-elastic-Bagnold). In the left panels (a,b,c) $f=34/89$ is used which
approximates $(3-\sqrt{5})/2$. In the right panels (d,e,f) data of $f$
which approximate $2-\sqrt{3}$ (d,e) and $(5-\sqrt{17})/2$ 
(f) are also included.
We have checked that finite size effects and commensurability effects
are not significant within the range of shear-rates used here.
In the semi-elastic-Bagnold model (f) the system size is limited
to avoid strong shear-banding effects.
The scaling functions of the granular systems (e,f)
are obtained by plotting $\tilde{\Sigma} \propto \Sigma/(\phi-\phi_{c})^{\beta'}$
vs $\tilde{\dot{\Gamma}} \propto \dot{\Gamma}/(\phi-\phi_{c})^{\Delta'}$
with $\phi_{c}=0.8415$ (random close packing density in two dimensions) and appropriate
exponents $\beta'$ and $\Delta'$.
}
\end{figure*} 

The current-voltage curves obtained at different values
of the coupling $\lambda$ are displayed in Fig.~3 (a).  
At stronger coupling $\lambda >1$ it appears that
a non-zero critical current $J_{c}(\lambda)= \lim_{E \to 0}
J(E,\lambda)$ exists, which becomes larger with increasing $\lambda$. 
This means that the Lorentz force does not
drive the vorticies significantly so that the 
system remains macroscopically superconducting
along the $y$ axis at strong enough coupling $\lambda$.
The finite critical current corresponds to the yield stress $\Sigma_{c}$ 
in rheology. The disordered configurations of vorticies shown in Fig.~2 suggests
that the system is an amorphous glassy state of vorticies.
On the other hand, at smaller coupling
$\lambda < 1$  and low enough $E$ the Ohm's law $J=\sigma(\lambda)E $ 
holds with finite linear conductivity $\sigma(\lambda)$ 
which becomes larger with increasing $\lambda$. Thus the vorticies can flow easily
producing significant energy dissipation at weak enough coupling $\lambda$.
At the isotropic point $\lambda=1$, 
we find a power law $J \propto E^{1-\alpha}$ with $1-\alpha=0.34(3)$.
This corresponds to the so called shear-thinning behaviour ($\alpha>0$) 
in rheology \cite{rheologybook}.

The above results strongly indicate that $\lambda_{c}=1$ is the critical 
point of a 2nd order phase transition at zero temperature.
This is supported by a good scaling collapse of the data onto a master curve 
as shown in Fig.~3 (d).
Our scaling ansatz is similar in spirit
to the ones used for the usual normal-to-superconducting
phase transition at finite temperatures \cite{WGI,FFH} which can also
be reinterpreted in the context of rheology \cite{Yoshino},
\beq
J=(\lambda-\lambda_{c})^{\beta} \tilde{J} 
\left(\frac{E}{(\lambda-\lambda_{c})^{\Delta}}\right)
\label{eq-scaling}
\eeq
The scaling function (master flow curve) 
is expected to behave asymptotically as
$\tilde{J}(x) \propto x$ for small enough $x$ in the Ohmic phase
($\lambda < \lambda_{c}$)
and $\lim_{x\to 0}\tilde{J}(x) \to {\rm const}$ in the superconducting
phase ($\lambda > \lambda_{c}$). The scaling ansatz \eq{eq-scaling} implies 
1) the linear conductivity diverges
as $\sigma(\lambda) \propto (\lambda-\lambda_{c})^{-(\Delta-\beta)}$ for
$\lambda \to \lambda_{c}^{-}$, 2) the critical current vanishes 
as $J_{c}(\lambda) \propto (\lambda-\lambda_{c})^{\beta}$ for
$\lambda \to \lambda_{c}^{+}$ and 
3) the critical behaviour 
$\tilde{J}(x) \propto x^{1-\alpha}=x^{\beta/\Delta}$
sets-in for large $x$.
Here we used the notations reflecting the analogy with the equilibrium critical
behaviour of ferro-magnets under magnetic field as noticed by Wolf, Gubser and
Imry \cite{WGI}: the shear plays the role of symmetry breaking field
like the magnetic field and the critical current emerges as an
order parameter like the magnetization (see \cite{Otsuki-Sasa} for a similar
argument in the context of rheology). As shown in Fig.~3, we find our scaling
ansatz works well with $\lambda_{c}=1$. 
We have checked that the universality does not depend on the use of different
irrational values of $f$ as demonstrated in Fig.~3 (d).

\begin{figure}[h]
\vspace*{.5cm}
\includegraphics[width=0.7\textwidth]{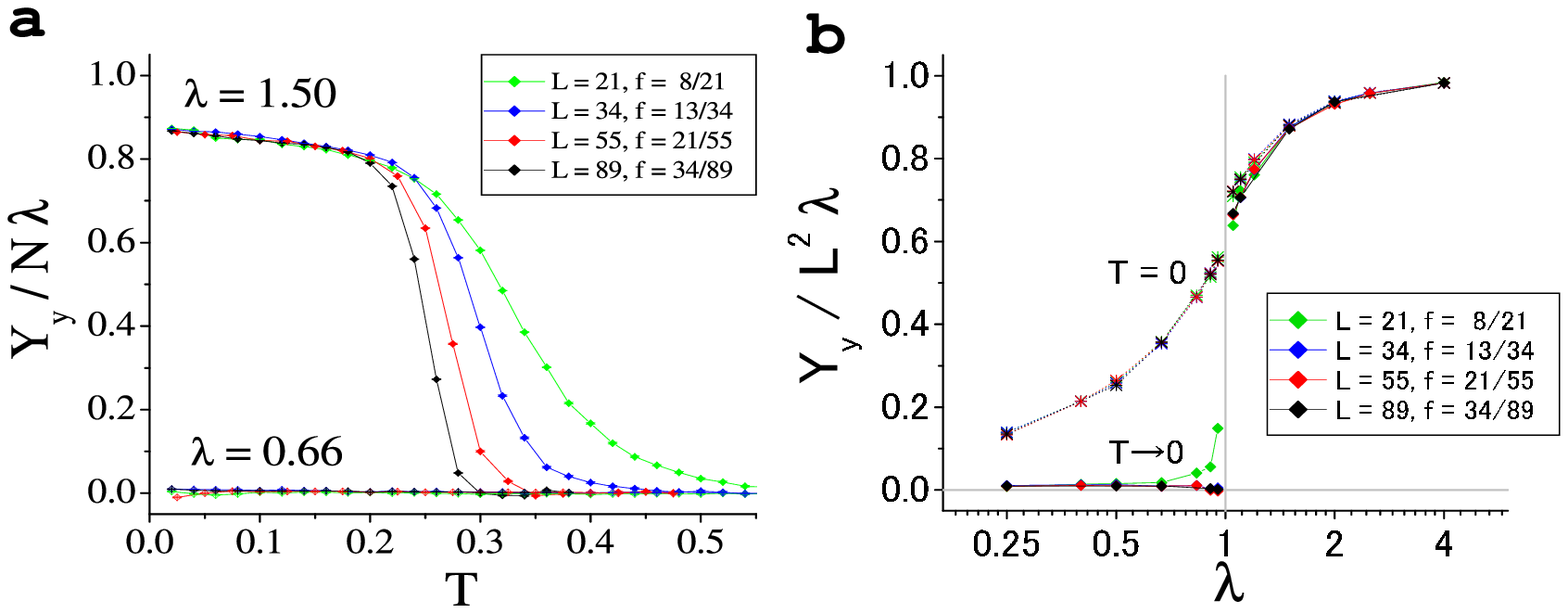}
\caption{ {\bf Static helicity (shear) modulus in the anisotropic JJA.} 
The helicity modulus $Y_{y}$ is obtained as equilibrium ensemble averages of 
the susceptibilities against twists along the $y$ axis.
(a) Temperature $T$ and size $L$ dependence of the helicity modulus $Y_{y}$ 
and (b) its extrapolation to $T\to 0$. In (b) helicity modulus at $T=0$
is also shown. 
The values of $f$ used here approximate $(3-\sqrt{5})/2$.
The periodic boundary condition is imposed along both $x$ and $y$
aixes on the system of size $N=L\times L$.
The standard simulated annealing method is used to generate 
the equilibrium ensemble (appendix \ref{appendix-static-response}).
}
\end{figure}

Let us note that the critial current discussed above
$J_{c}(\lambda)= \lim_{E \to 0} J(E,\lambda)$ corresponds to the {\it
dynamical yield stress} in rheology. On the other hand one can also
define the {\it quasi-static} critical current(s) needed to move out of a generic
metastable state (or the ground state), corresponding to the {\it static yield stress} in
rheology, as studied by Teitel and Jayaprakash \cite{Teitel-Jayaprakash} on
the JJA. In practice one can consider athermal quasi-static processes 
similar to those used in some recent studies on amorphous solids
\cite{anael,anael2,barrat}: starting from a metastable state reached from a
random initial configuration by a qunech, i.~e. deterministic energy descent process,
the system is subjected to externally induced uniform strain $\delta$
(See \eq{eq-mod-hamiltonian}) which is increased step by step. The
system relaxes down to an energy minimum by the energy descent process
after each small increment of $\delta$. As the result one finds that the
current $J(\delta)$ is a sawtooth-like function of $\delta$: piecewise linear lines corresponding to elastic
deformations broken by yield points at plastic events, as observed in amorphous solids \cite{anael,anael2,barrat}.
A critical current can be defined as the value of the current just
before reaching a yield point.
In \cite{Halsey} Halsey has found that typical value of such quasi-static
critical current $J_{c}^{\rm static}(1)$ is finite in the isotropic system $\lambda=1$. 
Moreover we found that $J_{c}^{\rm static}(\lambda)$ varies smoothly with $\lambda$
and remains finite even in the unjammed phase $\lambda < 1$ found
above \cite{comment-on-TJ-conjecture}. Our data of $J(E,\lambda)$ in the flow curves becomes smaller
than $J_{c}^{\rm static}(\lambda)$ meaning that the quasi-static current
needed to move out of a generic metastable state and dymamic critical currents
are distinct in the present system. Quite interestingly we observed that the flow curves 
become strongly dependent
on strain histories below $J_{c}^{\rm static}(\lambda)$. In practice we
had to use an annealing procedure to obtain stationary data: decrease
$E$ (shear rate) very slowly down to the target one. Slower shear rates
are needed to investigate smaller current $J$ (shear stress)
regions. Note that no annealing is performed in the athermal
quasi-static process discussed above. These observations suggest 
ruggedness of the energy landscape of the present system.

We also investigated static response to shear (See APPENDIX \ref{appendix-static-response} for the details.). 
As shown in Fig.~4, the static helicity (shear) modulus is very sensitive to the anisotropy. 
The figure shows that the helicity modulus remains zero down to $T \to 0$ 
for $\lambda < 1$ and becomes finite for $\lambda >1$. 
Thus at $T=0$ the isotropic point appears to be the critical point $\lambda_{c}=1$
being consistent with the dynamic response discussed so far.
Here let us recall again that just by symmetry, helicity modulus $Y_{x}/\lambda$ at anisotropy $\lambda$ 
is identical to $Y_{y}/\lambda'$ at $\lambda'=1/\lambda$. Then the fact that $\lambda_{c}=1$ means 
that this system is quite exotic: {\it fragile vortex matter} which behaves as a solid with 
respect to shear along one direction but liquid for the other direction.

Another remarkable feature is the difference of the static helicity modulus between $T=0$ and
$T\to 0$ limit which can be seen in Fig.~4 b).
The helicity modulus at $T=0$ only reflects local stability of energy minima.
This difference suggests the existence of certain softmodes with vanishingly 
small energy gaps as discussed in APPENDIX \ref{appendix-static-response}. 
This observation suggests that the vortex liquid behaviour at $T=0$ is realized by
some non-trivial softmodes due to frustrations.

Based on the above results we obtain the jamming phase diagram
of the anisotropic JJA as shown in Fig.~1 (e) which is surprisingly
similar to that of granular systems shown in Fig.~1 (f) 
\cite{Liu-Nagel,Jamming-Book,Nagel-group}.

Now let us turn to the semi-elastic model.
In Fig.~3 (b,c), we display the flow curves of the
semi-elastic model. The shear-stress due to
the inter-layer coupling terms also obeys the Newtonian or Bagnold 
scaling for small $\lambda$ at low enough  $\dot{\Gamma}$.
Flow curves at large $\lambda$ suggests existence of non-zero yield stresses
$\Sigma_{c}(\lambda) = \lim_{\dot{\Gamma}\to 0}\Sigma(\dot{\Gamma},\lambda)$.
We find again a power law behaviour $\Sigma \propto (\dot{\Gamma})^{1-\alpha}$
with $\alpha>0$ (shear-thinning) at $\lambda=1$. 
Indeed the scaling ansatz \eq{eq-scaling}
(with $E \to \dot{\Gamma}$ and $J \to \Sigma$)
works well again assuming $\lambda_{c}=1$ 
and the universality does not depend on the different irrational values of $f$ 
(Fig.~3 (e,f)).

Recent numerical simulations of granular materials 
with/without strong dissipation at the particle level 
(coefficient of restitution smaller than/equal to $1$) 
have found Bagnold/Newtonian scalings in the fluid phase and 
different critical exponents \cite{Xu-Hern,Hatano-Otsuki-Sasa,Olsson-Teitel,Hatano,Hatano-Yoshino}.
Quite interestingly the values of the shear-thinning exponent $1-\alpha$ 
found in our semi-elastic model with Bagnold/Newtonian frictions are 
$0.63$ and $0.42$ respectively in agreement with
the exponents of the corresponding two-dimensional granular systems
with repulsive linear spring forces between the particles.
For a comparison master flow curves 
of the granular systems \cite{Hatano-Yoshino}
are displayed in Fig.~3 (e,f).
Quite remarkably the functional forms of the master flow-curves 
themselves agree very well.

In the present paper we focused on the responses of the
systems to shear at zero temperature. 
Recent studies at the isotropic point $\lambda=1$ 
at finite temperatures suggest critical behaviour with $T \to 0$, i.~e.
$T_{c}(\lambda=1)=0$ with diverging length scales \cite{Park-Choi-Kim-Jeon-Chung,Granato,Granato2}. 
If this would be confirmed, our system would provide a fascinating example where both the 
(thermal) liquid-glass transition and the (athermal) unjamming-jamming transition
take place at the same thermodynamic point, $T=0$ and $\lambda_c=1$, demonstrating deep connection
between the two transitions. 
At least in this system, the jamming and glass transitions appears to be 
the two sides of a coin.  It would be interesting to further explore this connection
to make this statement more substantial.
A related interesting question is whether or not 
the jamming (glass) phase survives at finite temperatures. 
Then the possibilities are 
1) the jamming point at $\lambda=1$ at $T=0$ is an isolated critical
point or 2) a critical line $T_{c}(\lambda)$ starts from the jamming point
as shown in Fig.~1 (e). Our preliminary study points to the latter possibility.

We emphasize that the jamming transition here is purely due to
geometrical frustration which is free from any quenched disorder in
sharp contrast to the conventional vortex glasses \cite{FFH} for which
presence of random pinning centers are crucial. This is a much awaited,
concrete example of a jamming-glass transition purely due to geometrical
frustration \cite{Tarjus-review}. It is tempting to speculate that
similar phenomena may exist in frustrated magnets such as
antiferromagnets on triangular, kagome and pyrochlore lattices
\cite{pyrochlore}. 
We also note that it will be important and interesting to clarify how quenched disorders, 
which may not be completely avoided in experimental JJAs, come into play.

\vspace*{.2cm}
{\bf Acknowledgement} We thank T. Hatano, H. Hayakawa, 
H. Kawamura, K. Nemoto, H. Matsukawa, T. Ooshida, M. Otsuki, S. Sasa 
and S. Yukawa for stimulating discussions. 
The authors thank the Supercomputer Center, Institute for Solid State Physics, 
University of Tokyo for the use of the facilities.
This work is supported by Grant-in-Aid for Scientific Research
on Priority Areas "Novel States of Matter Induced by Frustration"
(1905200*) and by 21st Century COE program
¡ÈTopological Science and Technology¡É.

\appendix
\section{Rational approximants and commensurability effects}
\label{appendix-rational-approximants}

The purpose of the present paper is to analyzed physical properties
of the anisotropic JJA with irrrational vortex density $f$. However in order
to use the periodic boundary condition we had to use rational
approximants for a given irrational number. Here we explain how 
commensurability or matching effects emerge 
and explain how we avoid them in the present study.

For convenience we considered a family of irrational numbers 
called quadratic irrationals. A well known example is 
the golden mean $(1+\sqrt{5})/2$. Their rational approximants $f_{n}$ 
can be generated systematically by solving a simple recursion formulae,
\beq
f_{n}=\frac{a_{n-1}}{a_{n}} \qquad Aa_{n+1}=Ba_{n}+C a_{n-1}
\label{eq-gene-f}
\eeq
where $A$, $B$ and $C$ are integer coefficients and $a_{0}$, $a_{1}$
are certain integers. It is easy to verify that the $f_{n}$s converge
to an irrational number $f \equiv \lim_{n \to \infty} f_{n}$
which is a solution of a quadratic equation $Cf^{2}+Bf-A=0$.
For instance the golden mean $(1+\sqrt{5})/2$ can be obtained
using the Fibonacci numbers $a_{n}$ which satisfy the above recursion
formula with $A=B=C=1$ and $a_{2}=a_{1}=1$.
Examples of rational approximants $f_{n}$ are shown in Fig.~5 (a)
which approximate $(3-\sqrt{5})/2$.
They are generated by solving \eq{eq-gene-f} recursively with
$A=1$, $B=3$, $C=-1$ and $a_{2}=a_{1}=1$. It can be seen that
the approximation becomes better such that 
$|f-f_{n}|$ becomes smaller as $n$ increases.

In Fig.~5 (b) flow curves of the anisotropic JJA model at $T=0$ 
with the rational vortex densities $f_{n}$ are shown. 
Apparently flow curves of a given
anisotropy $\lambda$ converge to a limiting curve  
as $n$ is increased. We regard the latter as the flow curve of
$f=(3-\sqrt{5})/2$ at anisotropy $\lambda$. It can also be seen
that flow curves of a given approximant $f_{n}$ closely follow the limiting 
curve at large enough electric field $E$ (shear rate $\dot{\Gamma}$)
and deviate from it at lower $E$.
In the $n$ dpendent branches the current $J(E)$ (shear force $\Sigma$)
tends to saturate to some finite values in $E \to 0$ limit, i.~e. critical
current $J_{c}(\lambda,f_{n})$ (yield stress $\Sigma_{c}$) 
which decreases with increasing $n$ being consistent with the prediction by
Teitel and Jayaprakash \cite{Teitel-Jayaprakash} (see also \cite{halsey-staircase}).
The latter behaviour suggests 
that a periodic vortex lattice (crystal) \cite{Teitel-Jayaprakash,halsey-staircase,Mooji-group,gupta}
associated with a given
rational vortex density $f_{n}$ is formed and that the latter 
dictates physical properties of the system at length/time 
scales larger than its lattice spacing.
Thus we expect the so called 'Bingham fluid' behaviour (fluid with
finite yield stress) cannot be avoided for any small $\lambda$
for fixed $n$ and that the genuine fluid phase at $T=0$ 
is realized only for truly irrational vortex densities. 
On the other hand the above results shown
in Fig.~5 (b) suggest that physical properties at short enough
length/time scales, which are $n$ independent, reflect those
properties of irrational $f$.
Thus our strategy in the present work 
is to choose large enough $n$ such that the $n$ dependency
do not become relvant within the range of $E$ (shear rate $\dot{\Gamma}$) 
we choose to work on.

\begin{figure}[h]
\vspace*{.5cm}
\includegraphics[width=0.6\textwidth]{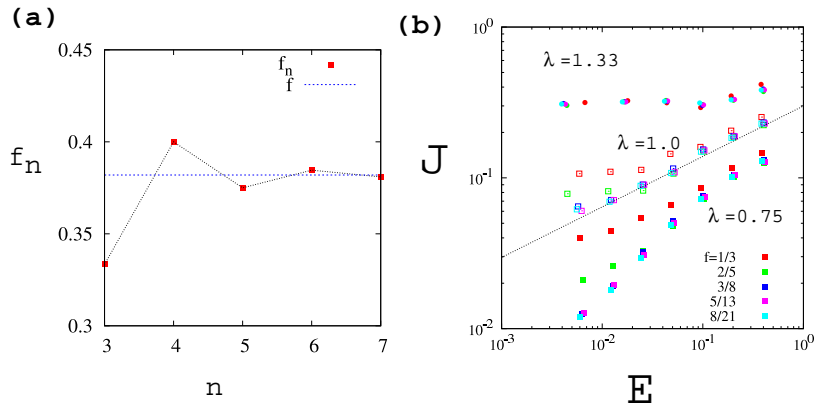}
\caption{{\bf Rational approximants 
and commensurability effects on the flow curves. }
(a) Approximants $f_{n}$ for $f=(3-\sqrt{5})/2$.
(b) Data of the flow curves of the anisotropic JJA at $T=0$ with
$f=1/3$ ($L=42$), $2/5$ ($40$), $3/8$($40$), $5/13$ ($39$) and
$8/21$ ($42$) are shown. The dotted straight line represents
the power law $J \propto E^{0.35}$ at the critical point. 
}
\end{figure}

\section{Static response with respect to shear}
\label{appendix-static-response}

To study static response to shear along, say $y$ axis, 
it is useful to consider a modified Hamiltonian,
\beq
H= -\sum_{<i,j> || e_{x}} \cos(\phi_{ij})
-\lambda \sum_{<i,j> || e_{y}} \cos(\phi_{ij}-\delta),  \qquad \phi_{ij} \equiv \theta_{i}-\theta_{j}-A_{ij}
\label{eq-mod-hamiltonian}
\eeq
with periodic boundary conditions along both $x$ and $y$ axes. Here
$e_{x}$ and $e_{y}$ are unit vectors along $x$ and $y$ axes. This is equivalent to consider
a system with twisted boundary condition (shear) with total phase difference $L \delta$ forced 
across the system along the $y$ axis \cite{Teitel-Jayaprakash}.

In equilbrium at temperature $T$ the free-energy $F(\delta)$ of the system under shear strain $\delta$ 
 can be defined.
Then the static helicity modulus $Y_{y}$ is defined as,
\beqa
Y_{y} \equiv \frac{\partial^{2} F}{\partial \delta^{2}} &&= \lambda \sum_{<i,j> || e_{y}} \langle \cos(\phi_{ij}) \rangle \nonumber\\
&& - \lambda \beta 
\sum_{<i,j>|| e_{y}} 
\sum_{<k,l>|| e_{y}} 
\left(
\langle \sin(\phi_{ij}) \sin(\phi_{kl})
\rangle - \langle \sin(\phi_{ij}) \rangle
\langle \sin(\phi_{kl}) \rangle
\right)
\label{eq-helicity-modulus}
\eeqa
where $\beta \equiv 1/(k_BT)$ and $\langle \ldots \rangle$ are equilibrium thermal averages at finite temperature $T$.
The 1st term on the r.~h.~s reflects direct elastic response around energy minima
with respect to a small externally induced shear strain. On the other hand, the 2nd term reflects relaxation
of the system against the external strain at finite temperatures.

Here the distinction between $T=0$ and $T \to 0$ is important. At $T=0$ only the 1st term
exists. However, the contribution of the 2nd term can remain, in principle, in $T \to 0$ limit
if the strength of the thermal fluctuation of the current $\sin(\phi_{ij})$ is $O(T)$.
Such a situation can arise if there are soft modes with vanishingly small energy gap
such that they remain thermally active at arbitrarily low temperatures.

To obtain the equilibrium ensemble to evaluate the helicity modulus at $\delta=0$,
we performed simulations of relaxational dynamics by numerically solving the Langevin equation 
\beq
 \frac{\partial \theta_{i}}{\partial t} 
= - \frac{\partial H}{\partial \theta_{i}}+\sqrt{2T}\xi_{i}(t)
\eeq 
with the Hamiltonian $H$ given in \eq{eq-jja-hamiltonian} 
and $\xi_{i}(t)$ being Gaussian noise with zero mean and unit variance, 
by the 2nd order Runge-Kutta method. The system is cooled with cooling rates 
$dT/dt=10^{-10}-10^{-9}$ starting from initial temperature 
at $T=0.3-1.0$. We have checked that the cooling rate is slow enough
by comparing with the results of $4$ times faster cooling rate.

%
%


\begin{thebibliography}{99}
\bibitem{comment-on-TJ-conjecture}
These observations suggest that the argument by Teitel-Jayaprakash
	\cite{Teitel-Jayaprakash} leading to the vanishing of the quasi-static critical current
	for irrational $f$ cannot be used always.
The assumption of smoothness of the variation of the current
with respect to the changes of the strain $\delta$, which underlies the use of the
mean-value theorem in \cite{Teitel-Jayaprakash}, becomes questionable if plastic deformations take place.
\bibitem{continuous-limt}
It is essential not to  lose 'granularity', which is the
source of frustration effects, by naively taking continuous limit $a \to 0$.
The irrational winding number $f$ can be identified with
the irrational number density $f$ of vorticies in a unit cell 
in the Josephson junction ladder.
\bibitem{runge-kutta}
We integrated \eq{eq-motion} with integration time step of $\delta t=0.1$
by the 4th order Runge-Kutta method starting from random initial conditions.
Initial $10^{5}$ steps are discarded and the time averages of the observables are 
taken in the following $10^{5}$ steps. We checked the latter 
protocol is sufficient to obtain stationary data within 
the shear rates and the system sizes used in the following analysis.
The results are averaged over $30-100$ independent runs with different random 
initial conditions.
\end{thebibliography}
\end{document}